\documentclass[aps,twocolumn,pra,showpacs,floatfix]{revtex4}

\usepackage{epsfig}
\usepackage{graphicx}
\usepackage{dcolumn}

\begin{document}
\title{Atomic calculations and search for variation of the fine
structure constant in quasar absorption spectra}
\author{V. A. Dzuba and V. V. Flambaum}
\address{School of Physics, University of New South Wales, Sydney 2052,
Australia}

\date{\today}

\begin{abstract}

A brief review of the search for variation of the fine structure constant
in quasar absorption spectra is presented.
Special consideration is given to the role of atomic calculations
in the analysis of the observed data.
A range of methods which allow to perform calculations for atoms
or ions with different electron structure and which cover practically
all periodic table of elements is discussed. Critical
compilation of the results of the calculations as well as a review
of the most recent results of the analysis are presented.

\end{abstract}

\maketitle

\section{Introduction}

Theories unifying gravity with other interactions as well as many
cosmological models allow for space-time variation of fundamental
constants. Experimental search for the manifestation of this
variation spans the whole lifetime of the Universe from
Big Bang nucleosynthesis to the present-day very precise
atomic clock experiments (see, e.g. reviews~\cite{Uzan,Flambaum07a}).
An evidence that the fine-structure constant $\alpha$ ($\alpha=e^2/\hbar c$)
might be smaller in early universe
has been found in the analysis of quasar absorption
spectra~\cite{Webb99,Webb01,Murphy01a,Murphy01b,Murphy01c,Murphy01d,Webb03}.
The analyzed data which came from the Keck telescope in Hawaii
included three independent samples containing 143
absorption systems which spread over red shift range $0.2 <z < 4.2$.
The fit of the data gives 
$\delta \alpha/\alpha=(-0.543 \pm 0.116) \times 10^{-5}$~\cite{murphy1}. 
If one assumes the linear dependence
of $\alpha$ on time, the fit of the data gives $d\ln{\alpha}/dt=
(6.40 \pm 1.35) \times 10^{-16}$ per year
(over time interval about 12 billion years).

A very extensive search for possible systematic errors~\cite{Webb03a} 
has shown that known systematic effects can not explain the result.
Although it is still not completely excluded that the effect may be
imitated by a large change of abundances of isotopes
in last 10 billion years, it would need a very unlikely
``conspiracy'' between several elements.
It had been checked that different isotopic abundances for any single
element can not imitate the observed effect. 

Recently two other groups of researchers~\cite{chand,Levshakov,Levshakov1}
applied the same method of the analysis to a different set of data obtained 
from the VLT telescope in Chile and reported no variation of $\alpha$.
There was an intensive debate in the literature about possible 
reasons for disagreement. It was argued in particular that
at least part of the disagreement may be attributed to the
spatial variation of $\alpha$. This argument is based on the
fact that Keck telescope is in the Northern hemisphere while VLT
telescope is in the Southern hemisphere.

The results of \cite{chand} were recently questioned in 
Refs.~\cite{murphy2,murphy3}. Re-analysis of the 
data of Ref.~\cite{chand} revealed flawed parameter estimation methods.
The authors of \cite{murphy2,murphy3} claim that the same spectral data fitted 
more accurately give  $\delta \alpha/\alpha=
(-0.64 \pm 0.36) \times 10^{-5}$ rather than  $ \delta \alpha/\alpha=
(-0.06 \pm 0.06) \times 10^{-5}$ as in Ref.~\cite{chand}. However, even this
new result may require further revision. 

All the results discussed above were obtained with the use of the 
so called many-multiplet (MM) method which was proposed in Ref.~\cite{MM}. 
This method uses atomic calculations
\cite{Dzuba99,Dzuba01,Dzuba02,Berengut04,Berengut05,Dzuba05,
Berengut06,Dzuba07,Savukov,Dzuba08}
to relate the change in the value 
of the fine structure constant to the change in the frequencies of
atomic transitions. It is more than an order of magnitude more
sensitive to the variation of the fine structure constant than
the analysis of the fine structure intervals used for this purpose
before.

In present paper we review the methods of atomic calculations 
used in the search for the variation of the fine structure constant
and present critical compilation of the most accurate results
of the calculations and discuss some future directions.

\section{Atomic calculations}

In atomic units $\alpha=1/c$, were $c$ is speed of light, and
$\alpha=0$ corresponds to non-relativistic limit.
Therefore, to reveal the dependence of atomic frequencies on $\alpha$ we
need to perform relativistic calculations based on the Dirac equation.
Doing this way we include leading relativistic corrections of the
order $(Z\alpha)^2$. The role of smaller corrections, such as Breit and 
quantum electrodynamic (QED) corrections will be discussed in 
section~\ref{sec:Breit}.

It is convenient to present the dependence of atomic frequencies on
the fine-structure constant $\alpha$ in the vicinity of its physical
value $\alpha_0$ in the form
\begin{equation}
  \omega(x) = \omega_0 + qx,
\label{omega}
\end{equation}
where $\omega_0$ is the present laboratory value of the frequency and
$x = (\alpha/\alpha_0)^2-1$, $q$ is the coefficient which is to be
found from atomic calculations. Note that
\begin{equation}
 q = \left .\frac{d\omega}{dx}\right|_{x=0}.
\label{qq}
\end{equation}
To calculate this derivative numerically we use
\begin{equation}
  q \approx  \frac{\omega(+\delta) - \omega(-\delta)}{2\delta}.
\label{deriv}
\end{equation}
and vary the value of $\alpha$ in the computer code.

We use relativistic Hartree-Fock method as a starting point of all
calculations.
The radial equation for single-electron
orbitals has the form (atomic units)
\begin{equation}
    \begin {array}{c} \frac{df_v}{dr}+\frac{\kappa_{v}}{r}f_v(r)-
    \left[2+\alpha^{2}(\epsilon_{v}-\hat{V}_{HF})\right]g_v(r)=0,  \\
    \frac{dg_v}{dr}-\frac{\kappa_{v}}{r}f_v(r)+(\epsilon_{v}-
    \hat{V}_{HF})f_v(r)=0, \end{array}
\label{Dirac}
\end{equation}
here $\kappa=(-1)^{l+j+1/2}(j+1/2)$, index $v$
replaces the three-number set of the principal quantum number, and
total and angular momentum: $n,j,l$.
$\hat{V}_{HF}$ is the Hartree-Fock potential.

Equations (\ref{Dirac}) with $\alpha=\alpha_0 \sqrt{\delta+1}$ are
solved self-consistently for all core states to find Hartree-Fock
potential of the atomic core. Then this potential is used to
calculate a full set of single-electron orbitals for the states 
above the core.

After single-electron states are calculated, the actual choice of the
methods to calculate many-electron states of valence electrons depends
on the number of valence electrons. 
Table~\ref{tb:methods} summarizes the methods used in the calculations. 
These methods will be discussed in more detail in following sections.

\begin{table}
\caption{{\em Ab initio} methods of atomic calculations
depending on the number of valence electrons ($N_v$).}
\label{tb:methods}
\begin{tabular}{lll}
\hline
$N_{v}$ & Method & Accuracy \\
\hline
1   & MBPT + all-order sums & 0.1 - 1\%   \\
2-8 & MBPT + Configuration interaction (CI+MBPT)& 1-10\%  \\
2-15 & Configuration interaction (CI) & 10-20\%  \\
\hline
\end{tabular}
\end{table}

\subsection{Atoms with one external electron}

Atoms and ions of astrophysical interest which can be considered as systems
with one external electron above closed shells in both ground and excited
states include C~IV, O~VI, Na~I, Mg~II, Al~III, Si~IV, Ca~II, Zn~II and Ge~II.
All calculations for these atoms and ions were performed in the $V^{N-1}$
approximation in which Eq.~(\ref{Dirac}) are solved for the closed-shell
core with the external electron removed. The states of valence electron
are calculated in the $V^{N-1}$ potential of the frozen core. Correlations
are also included using different numerical techniques. In most of the 
calculations (see, e.g. Ref.~\cite{Dzuba99}) the correlation potential method~\cite{CPM}
was used.

\begin{figure}
\includegraphics[scale=0.8]{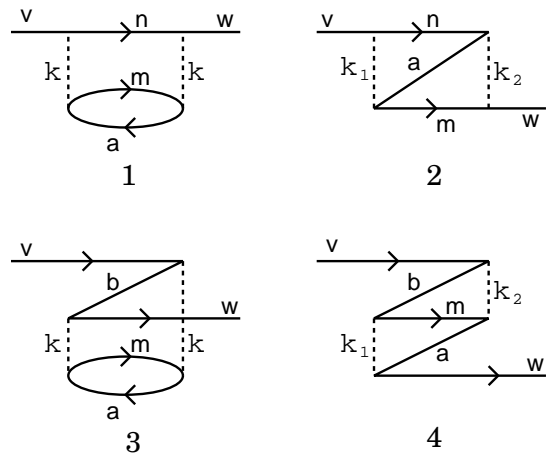}
\caption{Second-order diagrams for the matrix elements
$\langle v |\hat \Sigma^{(2)}_1| w \rangle$ of the single-electron 
correlation operator $\hat \Sigma^{(2)}_1$.}
\label{sigma1-fig}
\end{figure}

In this approach the correlation potential $\hat \Sigma$, which describes 
correlations between external and core electrons, is calculated in the 
lowest, second-order of the many-body perturbation theory (MBPT) using 
the B-spline basis set~\cite{Bsplines}. All four second-order diagrams 
for $\hat \Sigma$ are presented on Fig.~\ref{sigma1-fig}.

The $\hat \Sigma$ operator is a non-local operator similar to the
Hartree-Fock exchange potential. It is used in the equations for the states 
of an external electron to calculate the so-called Brueckner orbitals (BO). 
The orbitals and corresponding energies include all second-order correlations 
as well as higher-order correlations which correspond to  
$\langle \hat \Sigma \rangle^2$, $\langle \hat \Sigma \rangle^3$, etc.
The accuracy for the energies in this approach is usually few per cent or 
better.

Correlation potential method with the second-order $\hat \Sigma$ does
not include dominating higher-order correlations. The effect of the
higher-order correlations on the $q$-factors for mono-valent atoms of
astrophysical interest were studied in detail in Ref.\cite{Dzuba07}
within the single-double coupled cluster approach
and were found to be small.

\subsection{Atoms with open $s$ and $p$ shells}

Atoms or ions with two or more electrons on open $s$ and/or $p$ shells
are next in the complexity of the calculations. Systems found in
astrophysical observations include C~I, C~II, C~III, O~I, O~II,
O~III, O~IV, Mg~I and Al~II. 
We perform calculations for such systems with the use of the configuration
interaction method combined with the many-body perturbation theory
(CI+MBPT)~\cite{CI+MBPT}. Configuration interaction (CI) technique is used
for accurate treatment of the interaction between valence electrons
while the MBPT is used to include correlations between core and valence
electrons.
Similar to the case of atoms with one external electron, correlation
operator $\hat \Sigma$ is used to describe the core-valence correlations.
However, it now
consists of at least two parts. $\hat \Sigma_1$ describes correlations
between an external electron and the electrons in the core, while
$\hat \Sigma_2$ describes correlation correction (screening) to the 
Coulomb interaction between two external electrons caused by the core 
electrons. The $\hat \Sigma_1$ operator
is similar to the $\hat \Sigma$ operator used for atoms with one
external electron. The actual form for of the $\hat \Sigma$ operator 
depends on the choice of the potential in which core states are 
calculated. The simplest form is in the so-called $V^{N-M}$ 
approximation~\cite{VNM1,VNM2} which is a generalization of the
$V^{N-1}$ approximation used for atoms with one external electron.

In this approximation the $M$ external electrons are excluded from 
the initial Hartree-Fock procedure for the core to make sure
that the effective potential of the CI Hamiltonian for the valence 
states and the potential in which core electrons are calculated
are the same. This is the key for the simplest form of the MBPT.
This approach gives good results for atoms with open $s$ or $p$ 
shells because electrons of these shells are localized on large
distances and have very small effect on the wave functions of
the core states (in spine of large effect on their energies, 
see Refs.~\cite{VNM1,VNM2} for details).

\begin{figure}
\includegraphics[scale=0.6]{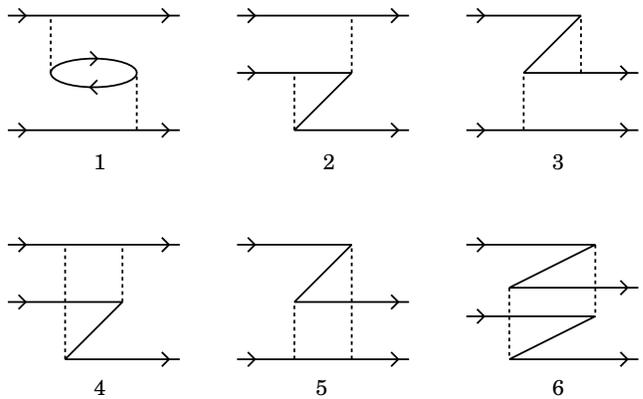}
\caption{Second-order diagrams for double-electron correlation
operator $\hat \Sigma^{(2)}_2$.}
\label{sigma2-fig}
\end{figure}

In the $V^{N-M}$ approximation the $\hat \Sigma_1$ operator
is identical to the $\hat \Sigma$ operator used for atoms with one
external electron. Corresponding diagrams are presented 
on Fig.~\ref{sigma1-fig}. Diagrams for $\hat \Sigma_2$
are presented on Fig.~\ref{sigma2-fig}.

If the potential in which core states are calculated is different from the
$V^{N-M}$ potential, then both  $\hat \Sigma_1$ and  $\hat \Sigma_2$
include the so-called {\em subtraction} diagrams which account for
this difference (see Ref.~\cite{CI+MBPT} for details).

The effective CI Hamiltonian for the valence electrons has the
form
\begin{equation}
  \hat H^{CI} = \sum_{i=1}^M \left[ \hat h_{0i} + \hat \Sigma_{1i} \right]
+ \sum_{i<j}^M \left[ \frac{1}{r_{ij}} + \hat \Sigma_{2ij} \right],
\label{HCI}
\end{equation}
where $\hat h_0$ is a single-electron Hartree-Fock Hamiltonian which
corresponds to Eq.~(\ref{Dirac}).

The CI equations
\begin{equation}
  \hat H^{CI} \Psi(r_1, \dots ,r_M) = E \Psi(r_1, \dots ,r_M)
\label{CI}
\end{equation}
are solved by matrix diagonalization using the the many-electron
basis states constructed from the B-splines. Therefore, 
the B-spline basis set serves the dual purpose in the calculations:
to calculate the correlation operator $\hat \Sigma$ and to solve
the CI equations.

The details of the calculations vary slightly from atom to atom. 
For example,  $\hat \Sigma$ was not included for light atoms like
carbon. Here relativistic corrections are small and high accuracy
of the calculations is not needed.

\subsection{Atoms with open $d$ or $f$ shells}

Atoms with open $d$ and $f$ shells are most difficult for calculations.
This is mostly due to large number of electrons which enter the
CI calculations. This makes it practically impossible to saturate
the basis with an arbitrary choice of single-electron basis
states.

Also, accurate treatment of the core-valence correlations is much
more difficult. First, the $V^{N-M}$ approximation is not valid 
because $d$ and $f$ electrons are inside the core and cannot be
neglected in the initial HF calculations for the core. 
Therefore, subtraction diagrams must be included. 
Second, for atoms with more than two valence
electrons, a new kind of $\hat \Sigma$ operator appear even
in the lowest second-order of the MBPT, the three-electron
operator $\hat \Sigma_3$ (see Ref.~\cite{CI+MBPT} for details).
The most detailed study of an atom with open $d$-shell, which
used large basis, all three $\hat \Sigma$-operators
$\hat \Sigma_1$,$\hat \Sigma_2$ and $\hat \Sigma_3$ and Breit
interaction were performed in Ref.~\cite{Porsev} for the Fe~II ion.

We use a less sophisticated approach which still gives
reasonably accurate results~\cite{Dzuba99,Dzuba01,Dzuba02,Berengut04,Dzuba08}.
It is based on the CI technique with simplified treatment of the
core-valence correlations (see below) and different choice of the
basis in different calculations for the valent states.
In our early calculations~\cite{Dzuba99,Dzuba01,Dzuba02,Berengut04} we
used either a B-spline basis set or a set of orbitals which was
constructed using a recurrent procedure suggested in Ref.\cite{Bogdanovich}.
In this procedure an virtual orbital is constructed from a lower one
by multiplying it by distance $r$ and orthogonalizing it to all lower
orbitals (see also Ref.~\cite{CI+MBPT}).

In the most recent work~\cite{Dzuba08} we used a more sophisticated approach
in which a set of non-orthogonal Hartree-Fock valence states was
constructed.
The self-consistent HF procedure was done for each configuration of 
interest separately. Only one single-electron basis function in each
partial wave is included in the CI for each configuration. However,
this functions represent good initial approximations since they come
from the self-consistent HF calculations.

To include core-valence correlations $\hat \Sigma_1$ is replaced by
a correction to the local part of the HF potential of the atomic core
\begin{equation}
  \delta V = - \frac{\alpha_p}{2(r^4+a^4)}.
\label{dV}
\end{equation}
Here $\alpha_p$ is polarization of the core and $a$ is a cut-off parameter
(we use $a = a_B$).
The form of the $\delta V$ is chosen to coincide with the standard polarization
potential on large distances ($-\alpha_p/2r^4$). The $\alpha_p$ is treated as a 
fitting parameter and values of $\alpha_p$ for each configuration are chosen
to reproduce their position in experimental spectrum.
The $\hat \Sigma_2$ and $\hat \Sigma_3$ operators are either neglected or
simulated by introducing screening parameters for Coulomb integrals.
It is usually assumed that the screening parameters $f_k$ depend only
on the multipolarity $k$ of the Coulomb interaction and their values 
are chosen to have better agreement with the experimental energies.
It is also assumed that all fitting parameters $\alpha_p$ and $f_k$ 
do not depend on $\alpha$. This is justified because the fitting represents
only small correction to the energies.

The effective CI Hamiltonian has the form
\begin{equation}
  \hat H^{CI} = \sum_{i=1}^M \left[ \hat h_{0i} + \delta V_{i} \right]
+ \sum_{i<j}^M \frac{1}{r_{ij}}.
\label{HCId}
\end{equation}
It can be obtained from (\ref{HCI}) by dropping $\hat \Sigma_2$ and
replacing $\hat \Sigma_1$ by $\delta V$ [see Eq.~(\ref{dV})].

\subsection{Level pseudo-crossing}

Calculations may be complicated significantly by the phenomenon
of level pseudo-crossing~\cite{Dzuba01,Dzuba02}. In atoms
with dense spectrum the states of the same total momentum $J$ and 
parity and separated by small energy interval may be strongly
mixed which in turn may lead to instability of the calculations
of the $q$ coefficients. This may be considered as level pseudo-crossing
in the vicinity of the physical value of $\alpha$ if energies of
these two states are considered as functions of $\alpha^2$.
In two-level approximation the $q$ coefficients for each state
are
\begin{equation}
\begin {array}{c}
  q_1 = \cos^2 \phi \ q_1^{(LS)} + \sin^2 \phi \ q_2^{(LS)}, \\
  q_2 = \sin^2 \phi \ q_1^{(LS)} + \cos^2 \phi \ q_2^{(LS)},
\end{array}
\label{q12}
\end{equation}
where $\phi$ is the mixing angle and $q_1^{(LS)}$ and $q_2^{(LS)}$
are the $q$ coefficients for the states which correspond to the
values of $\alpha$ far from the crossing. Very often atomic
states in the absence of crossing are well described by the $LS$
coupling scheme which is indicated by the $(LS)$ superscripts in Eq.~\ref{q12}.
If $q_1^{(LS)}$ and $q_2^{(LS)}$ are significantly different than
the values of $q_1$ and $q_2$ strongly depend on the mixing angle $\phi$
which leads to instability of the results of calculations.

The best way of dealing with this instability is to use experimental
values of the Land\'{e} $g$-factors. In the two level approximation 
the $g$-factors of two states are given by the formula very similar to 
Eq.~(\ref{q12}): 
\begin{equation}
\begin {array}{c}
  g_1 = \cos^2 \phi \ g_1^{(LS)} + \sin^2 \phi \ g_2^{(LS)}, \\
  g_2 = \sin^2 \phi \ g_1^{(LS)} + \cos^2 \phi \ g_2^{(LS)},
\end{array}
\label{g12}
\end{equation}
with the same mixing angle $\phi$ as in (\ref{q12}) and with 
$g_1^{(LS)}$ and $g_2^{(LS)}$ given by
\begin{equation}
  g^{(LS)} = 1 + \frac{J(J+1)-L(L+1)-S(S+1)}{2J(J+1)},
\label{gnr}
\end{equation}
where $J,L,S$ are total and angular momentums and spin
of the states. Eqs.~(\ref{g12},\ref{gnr}) can be used to find the mixing
angle $\phi$ which than is used to correct the calculated values of $q_1$
and $q_2$.

Note that it follows from (\ref{q12}) that the sum $g_1+g_2 =
g_1^{(LS)} + g_2^{(LS)}$ and does not depend on $\phi$. This can
be used to check the accuracy of the two-level approximation.

This scheme cannot be used if experimental values of $g$-factors
are not known or if $g_1+g_2 \neq g_1^{(LS)} + g_2^{(LS)}$ or
if $g_1^{(LS)} = g_2^{(LS)}$. If experimental $g$-factors are
known but $g_1+g_2 \neq g_1^{(LS)} + g_2^{(LS)}$ then a
multilevel consideration may help as explained in Ref.~\cite{Dzuba02}.
Otherwise the only way to get reliable results is to fit the energies
to experimental values to the accuracy sufficiently better than
the small energy interval between strongly mixed states (see, e.g.~\cite{Dzuba08}).

\subsection{Breit and QED corrections}
\label{sec:Breit}

So far we have considered relativistic corrections to atomic energies
which come from the Dirac equation. These corrections have an order
of $(Z\alpha)^2$. There are however other smaller corrections like
Breit and quantum electrodynamic corrections. It is important to
understand the role these corrections may play in the search for
variation of $\alpha$. Breit corrections are proportional to
$\alpha^2$ but smaller than 2 power of nuclear charge $Z$. Its
relative contribution is therefore larger for light atoms. 
In contrast, the QED corrections are proportional to $\alpha^3$ but
stronger than $Z^2$ functions of $Z$. Their relative contribution is
larger for heavy atoms. The role of the Breit interaction was studied
in detail in Ref.~\cite{Savukov} for mono and double-valent electron
atoms and ions of astrophysical interest and in Ref.~\cite{Porsev}
for Fe~II. The role of the QED correction has not been studied yet 
but expected to be small.

The following form of the Breit operator is used in the
relativistic calculations for many-electron atoms (atomic units)
\begin{equation}
    \hat H^B = - \frac{{\hat{\bf \alpha}_1}\cdot{\hat{\bf \alpha}_2}+
    ({\hat{\bf \alpha}_1}\cdot{\bf \hat{n}})({\hat{\bf \alpha}_2}
    \cdot{\bf \hat{n}})}{2r}.
\label{HBreit}
\end{equation}
Here ${\bf r} = {\bf \hat{n}}r$, $r$ is the distance between
electrons and $\hat{\bf \alpha}_i$ is the $\alpha$-matrix of the
corresponding electrons. This is a low frequency limit of the
relativistic correction to the Coulomb interaction between
electrons. It contains magnetic interaction and retardation.

It is important to include $\hat H^B$ into Hartree-Fock Hamiltonian
to take into account the effect of Breit interaction on the self-consistent
field (relaxation effect). for atoms with two or more external electrons
Breit term is also included as a correction to the 
Coulomb interaction in the CI Hamiltonian.

It turns out~\cite{Savukov,Porsev} that Breit corrections to the
$q$-values are relatively small and unlikely affect the analysis
of quasar spectra in terms of variation of $\alpha$. However,
it is useful to include them for more accurate results.
For example, as it was demonstrated in Ref.~\cite{Savukov}
inclusion of Breit interaction significantly improves the
agreement between theoretical and experimental fine structure
making the result to be more accurate and therefore more reliable.

Similar role is expected for the QED corrections. But it is the
subject of future work. An adequate method to include leading
QED corrections to the energies of many-electron atoms has been
developed in Ref.~\cite{Ginges}.

\section{Results of calculations}

\begin{table}
\begin{center}
\caption{Energies of the transitions from the ground
state of single-valent atoms and ions of astrophysical interest and
corresponding relativistic $q$ coefficients (cm$^{-1}$).}
\label{tb:one}
\begin{tabular}{rllrrr}
$Z$ & \multicolumn{1}{c}{Atom} & \multicolumn{1}{c}{State} & E(expt)
& E(calc) & 
\multicolumn{1}{c}{$q$} \\
\hline
 6  & C~IV   & $2p_{1/2}$ & 64484 & 64504 &  115(2) \\
    &        & $2p_{3/2}$ & 64591 & 64636 &  222(2) \\
 11 & Na~I   & $3p_{1/2}$ & 16956 & 16961 &   45(0) \\
    &        & $3p_{3/2}$ & 16973 & 16979 &   62(0) \\
    &        & $4p_{1/2}$ & 30267 & 30066 &   57(1) \\
    &        & $4p_{3/2}$ & 30273 & 30072 &   51(1) \\
 12 & Mg~II  & $3p_{1/2}$ & 35669 & 35687 &  121(1) \\
    &        & $3p_{3/2}$ & 35761 & 35784 &  212(1) \\
    &        & $4p_{1/2}$ & 80620 & 80463 &  161(1) \\
    &        & $4p_{3/2}$ & 80650 & 80496 &  192(1) \\
 13 & Al~III & $3p_{1/2}$ & 53684 & 53723 &  224(1) \\
    &        & $3p_{3/2}$ & 53917 & 53970 &  458(2) \\
    &        & $4p_{1/2}$ &143633 &143538 &  337(2) \\
    &        & $4p_{3/2}$ &143714 &143623 &  417(3) \\
 14 & Si~IV  & $3p_{1/2}$ & 71290 & 71352 &  361(2) \\
    &        & $3p_{3/2}$ & 71750 & 71836 &  823(2) \\
    &        & $4p_{1/2}$ &218267 &218226 &  597(4) \\
    &        & $4p_{3/2}$ &218429 &218397 &  760(4) \\
 20 & Ca~II  & $4p_{1/2}$ & 25192 & 25086 &  222(1) \\
    &        & $4p_{3/2}$ & 25414 & 25315 &  446(3) \\
 30 & Zn~II  & $4p_{1/2}$ & 48481 & 48721 & 1541(7)   \\
    &        & $4p_{3/2}$ & 49355 & 49606 & 2452(13)  \\
\hline
\end{tabular}
\end{center}
\end{table}

\begin{table*}
\begin{center}
\caption{Energies of the transitions from the ground
state of Fe~I and Fe~II and
corresponding relativistic $q$ coefficients (cm$^{-1}$).}
\label{tb:fe}
\begin{tabular}{llllllcccrr}
\multicolumn{1}{c}{Atom} & \multicolumn{3}{c}{Ground state} & \multicolumn{3}{c}{Upper state} & 
\multicolumn{2}{c}{Energy} & 
\multicolumn{2}{c}{$q$} \\
 or ion & 
\multicolumn{1}{c}{Config.} &\multicolumn{1}{c}{Term} &\multicolumn{1}{c}{$J$} &  
\multicolumn{1}{c}{Config.} &\multicolumn{1}{c}{Term} &\multicolumn{1}{c}{$J$} &  
\multicolumn{1}{c}{Expt.\cite{NIST}} & \multicolumn{1}{c}{Calc.} & 
\multicolumn{1}{c}{\cite{Dzuba08}} & \multicolumn{1}{c}{\cite{Porsev}}\\
\hline
Fe~I & $3d^64s^2$ & $^5$D & 4  & $3d^6 4s4p$   & $^5$D$^o$ & 4 & 25899 & 26428 &   999(300) & \\
     &            &       &    & $3d^6 4s4p$   & $^5$F$^o$ & 5 & 26874 & 27432 &   880(260) & \\
     &            &       &    & $3d^6 4s4p$   & $^5$P$^o$ & 3 & 29056 & 29340 &   859(260) & \\
     &            &       &    & $3d^5 4s^24p$ & $^5$D$^o$ & 4 & 33095 & 32680 &  2494(750) & \\
     &            &       &    & $3d^5 4s^24p$ & $^5$D$^o$ & 3 & 33507 & 33134 &  3019(900) & \\
     &            &       &    & $3d^5 4s^24p$ & $^5$F$^o$ & 5 & 33695 & 32522 &  2672(800) & \\
     &            &       &    & $3d^5 4s^24p$ & $^5$D$^o$ & 4 & 39625 & 39544 &  1680(500) & \\
     &            &       &    & $3d^5 4s^24p$ & $^5$F$^o$ & 5 & 40257 & 40194 &  1042(300) & \\

Fe~II & $3d^64s$ & $^6$D & 9/2 & $3d^64p$ & $^6$D$^o$ & 9/2  & 38458 & 38352 & 1330(150) & 1410(60)\\
      &          &       &     & $3d^64p$ & $^6$D$^o$ & 7/2  & 38660 & 38554 & 1490(150) & 1540(40) \\
      &          &       &     & $3d^64p$ & $^6$F$^o$ & 11/2 & 41968 & 41864 & 1460(150) & 1550(60) \\
      &          &       &     & $3d^64p$ & $^6$F$^o$ & 9/2  & 42114 & 42012 & 1590(150) & 1660(60) \\
      &          &       &     & $3d^64p$ & $^6$P$^o$ & 7/2  & 42658 & 42715 & 1210(150) & 1540(400) \\
      &          &       &     & $3d^64p$ & $^4$F$^o$ & 7/2  & 62065 & 65528 & 1100(300) & 1560(500) \\
      &          &       &     & $3d^54s4p$ & $^6$P$^o$ & 7/2 &62171 & 65750 &-1300(300) &-1030(300) \\
\hline
\end{tabular}
\end{center}
\end{table*}

\begin{table*}
\begin{center}
\caption{Energies of the transitions from the ground
state of atoms and ions of astrophysical interest and
corresponding relativistic $q$ coefficients (cm$^{-1}$).}
\label{tb:other}
\begin{tabular}{rlllllrrr}
$Z$ & \multicolumn{1}{c}{Atom} & \multicolumn{2}{c}{Ground} &  \multicolumn{2}{c}{Upper} & 
 \multicolumn{2}{c}{Energy} & 
\multicolumn{1}{c}{$q$} \\
 & or ion & \multicolumn{2}{c}{state} & \multicolumn{2}{c}{state} &
\multicolumn{1}{c}{Expt.\cite{NIST}} & \multicolumn{1}{c}{Calc.} & \\
\hline
 6 & C~I    & $2s^22p^2$ & $^3$P$_0$ & $2s2p^3$ & $^3$D$^o_3$     & 64087 & 66722 & 151(60) \\
   &        &            &           & $2s2p^3$ & $^3$D$^o_1$     & 64090 & 66712 & 141(60) \\
   &        &            &           & $2s2p^3$ & $^3$D$^o_2$     & 64091 & 66716 & 145(60) \\
   &        &            &           & $2s2p^3$ & $^3$P$^o_1$     & 75254 & 75978 & 111(60) \\
   &        &            &           & $2s2p^3$ & $^3$S$^o_1$     &105799 &100170 & 130(60) \\

 6 & C~II   & $2s^22p$ & $^2$P$^o_{1/2}$ & $2s^22p$ & $^2$P$^o_{3/2}$ &    63 &    74 &  63(1) \\
   &        &          &                 & $2s2p^2$ & $^2$D$_{5/2}$   & 74930 & 76506 & 179(20) \\
   &        &          &                 & $2s2p^2$ & $^2$D$_{3/2}$   & 74933 & 76503 & 176(20) \\
   &        &          &                 & $2s2p^2$ & $^2$S$_{1/2}$   & 96494 & 97993 & 161(30) \\
 6 & C~III  & $2s^2$   & $^1$S$_0$       & $2s2p$   & $^1$P$^o_1$     &102352 &103955 & 163(1) \\

 7 & N~V    & $2s$     & $^2$S$_{1/2}$   & $2p$     & $^2$P$^o_{3/2}$ & 80722 & 81607 & 492(50) \\
 8 & O~II   & $2s^22p^3$ & $^4$S$^o_{3/2}$& $2s2p^4$ & $^4$P$_{5/2}$   &119873 &122620 &   346(50) \\
   &        &            &                & $2s2p^4$ & $^4$P$_{3/2}$   &120000 &122763 &   489(50) \\
   &        &            &                & $2s2p^4$ & $^4$P$_{1/2}$   &120083 &122848 &   574(50) \\

 8 & O~III  & $2s^22p^2$ & $^3$P$_0$   & $2s2p^3$ & $^3$D$^o_1$   &120058 &121299 &   723(50) \\
   &        &            &             & $2s2p^3$ & $^3$P$^o_2$   &142382 &143483 &   726(50) \\

 8 & O~IV   & $2s^22p$ & $^2$P$^o_{1/2}$ & $2s2p^2$ & $^2$D$_{3/2}$ &126950 &129206 &   840(50) \\

 8 & O~VI   & $2s$     & $^2$S$_{1/2}$ & $2p$     & $^2$P$^o_{1/2}$ & 96375 & 96501 &   309(50) \\
   &        &          &               & $2p$     & $^2$P$^o_{3/2}$ & 96908 & 97091 &   913(50) \\

12 & Mg~I   & $3s^2$  & $^1$S$_0$      & $3s3p$   & $^1$P$^o_1$     & 35051 & 35050  &   85(1) \\
   &        &         &                & $3s4p$   & $^1$P$^o_1$     & 49347 & 49277  &   80(1) \\
13 & Al~II  & $3s^2$  & $^1$S$_0$      & $3s3p$   & $^1$P$^o_1$     & 59852 & 59800  &  270(1) \\

14 & Si~II  & $3s^23p$ & $^2$P$^o_{1/2}$ & $3s3p^2$ & $^2$D$_{3/2}$ & 55309 & 54655 &   520(30) \\
   &        &          &                 & $3s^24s$ & $^2$S$_{1/2}$ & 65500 & 54675 &   50(30) \\

22 & Ti~II  & $3d^24s$ & $^4$F$_{3/2}$ & $3d^24p$ & $^4$G$^o_{5/2}$ & 29544 & 28097 &   396(50) \\
   &        &          &               & $3d^24p$ & $^4$F$^o_{3/2}$ & 30837 & 29401 &   541(50) \\
   &        &          &               & $3d^24p$ & $^4$F$^o_{5/2}$ & 30959 & 29521 &   673(50) \\
   &        &          &               & $3d^24p$ & $^4$D$^o_{1/2}$ & 32532 & 31143 &   677(50) \\
   &        &          &               & $3d^24p$ & $^4$D$^o_{3/2}$ & 32603 & 31227 &   791(50) \\
   &        &          &               & $3d4s4p$ & $^4$D$^o_{1/2}$ & 52339 & 50889 & -1564(150) \\
   &        &          &               & $3d4s4p$ & $^4$F$^o_{3/2}$ & 52330 & 51341 & -1783(150) \\

22 & Ti~III & $3d^2$   & $^3$F$_2$     & $3d4p$   & $^3$D$^o_1$     & 77000 & 80558 & -1644(150) \\

24 & Cr~II  & $3d^5$ & $^6$S$_{5/2}$ & $3d^44p$ & $^6$P$^o_{3/2}$ & 48398 & 48684 & -1360(150) \\
   &        &        &               & $3d^44p$ & $^6$P$^o_{5/2}$ & 48491 & 48790 & -1280(150) \\
   &        &        &               & $3d^44p$ & $^6$P$^o_{7/2}$ & 48632 & 48947 & -1110(150) \\

25 & Mn~II  & $3d^54s$ & $^7$S$_3$   & $3d^54p$ & $^7$P$^o_{2}$   & 38366 & 38424 &   869(150) \\
   &        &          &             & $3d^54p$ & $^7$P$^o_{3}$   & 38543 & 38585 &  1030(150) \\
   &        &          &             & $3d^54p$ & $^7$P$^o_{4}$   & 38807 & 38814 &  1276(150) \\
   &        &          &             & $3d^44s4p$ & $^7$P$^o_{2}$ & 83256 & 83363 & -3033(450) \\
   &        &          &             & $3d^44s4p$ & $^7$P$^o_{3}$ & 83376 & 83559 & -2825(450) \\
   &        &          &             & $3d^44s4p$ & $^7$P$^o_{4}$ & 83529 & 83818 & -2556(450) \\

28 & Ni~II  & $3d^9$ & $^2$D$_{5/2}$ & $3d^84p$ & $^2$F$^o_{7/2}$ & 57080 & 56067 & -700(250) \\
   &        &        &               & $3d^84p$ & $^2$D$^o_{5/2}$ & 57420 & 56520 &-1400(250) \\

32 & Ge~II  & $4s^24p$ & $^3$P$^o_{1/2}$ & $4s^25s$ & $^2$S$_{1/2}$   & 62402 & 62870 & -630(40) \\
\hline
\end{tabular}
\end{center}
\end{table*}

The results of the calculated $q$-values for lines which were
observed in quasar absorption spectra are presented in Tables~\ref{tb:one},
\ref{tb:fe} and \ref{tb:other}. Calculated and experimental energies 
are also presented to illustrate the accuracy of calculations. 
Experimental energies are taken from the NIST website~\cite{NIST}.
Estimations of uncertainties
for the values of $q$-factors are based on the sensitivity of the results
to variations of the calculation scheme and on the comparison of the
experimental and theoretical energies and fine structure intervals.
In cases when several calculations are available only latest most accurate 
results are presented. For the case of Fe~II we also present the results 
of independent calculations by the St. Petersburg group~\cite{Porsev}.

The analysis of the values of the $q$-factors reveal an interesting 
picture. These values vary strongly from atom to atom and from 
one transition to another. The values can be very small or can be
large positive or large negative. This is due to the effect of several
factors which can be qualitatively illustrated by the value
of the relativistic energy shift of a single-electron state~\cite{Dzuba99}
\begin{equation}
  \Delta_n = \frac{E_n}{\nu} (Z\alpha)^2 \left[ \frac{1}{j+1}-C(Z,j,l)\right],
\label{deltan}
\end{equation}
where $E_n$ is the energy of the single-electron state $n$,
$\nu$ is the effective principal quantum number ($\nu = 1/\sqrt{-2E_n}$),
$j$ is the total momentum of the state and constant $C$ ($C \approx 0.6$) 
simulates the effect of exchange interaction and other many-body effects.
The value of the $q$-factor for a particular atomic transition can be 
roughly estimated assuming that the transition involves the change of
a state of just one electron. Then
\begin{equation}
\centering
  q \approx \Delta_n - \Delta_{n'},
\label{ddn}
\end{equation}
where state $n'$ is above state $n$.

The first and most obvious feature of the $q$-factors is its $Z^2$
dependence on the nuclear charge $Z$. However, it also depends, e. g.
on the ionization degree. This is because the absolute value of the
single electron energy $E_n$ (\ref{deltan}) is larger for ions.
Also, dependence of the energy shift (\ref{deltan}) on the total
electron momentum $j$ suggests that the value of $q$ is likely
to by positive for the $s-p$ transitions and negative for the 
$p-d$ transitions.

All these features of the $q$-factor values are reproduced in the 
accurate numerical calculations (see Tables \ref{tb:one},\ref{tb:fe} 
and \ref{tb:other}). This complex dependence of atomic frequencies
on $\alpha$ is very important for study of possible systematics.
It is very unlikely that any other unknown effect exhibits
exactly the same features. Therefore, if the results of the
analysis based on different transitions give the same result
for variation of $\alpha$ then the mimic of the effect by any
systematic is very improbable.

Theoretic uncertainty for the values of $q$ varies significantly
from atom to atom. The most accurate results are for atoms which can
be considered as atoms with single valence electron above closed
shells in both ground and excited states. The results for such
atoms are presented in Table~\ref{tb:one}. They were obtained
using second-order MBPT~\cite{Savukov} as well as the single-double
coupled cluster approximation combined with the third-order 
MBPT~\cite{Dzuba07}. Small uncertainty is due to excellent agreement
between different approaches and between experimental and
theoretical data for the energies and fine structure intervals.
Breit interaction was also included in Ref.~\cite{Savukov}.
Although Breit contributions are very small for the purposes
of the analysis they are larger than the uncertainty of
the calculation of the correlations. It was demonstrated
in Ref.~\cite{Savukov} that inclusion of Breit contributions 
improve significantly theoretical fine structure brining it
to almost perfect agreement with the experiment. 

Table \ref{tb:fe} presents results for Fe~I and Fe~II. Iron is one of the
most important elements used in the analysis. It is used more often than
most of other elements and it is also most studied theoretically.
First analysis of the quasar absorption spectra compared shift
of frequencies of Fe~II to those of Mg~I and Mg~II. The values
of the $q$ coefficients for magnesium are small compared to those
for Fe~II. Therefore, one can say they were used as ``anchors''
against which the shift of Fe~II lines was measured.
When it was realized that Fe~II has also large negative shifters
(see last line of Table~\ref{tb:fe}) it was suggested that Fe~II
alone can be used in the analysis by comparing lines with
positive and negative values of $q$. This may help to eliminate
certain types of systematic errors~\cite{Porsev}.

Calculations for Fe~II were carried out by means of the CI method
in Refs.~\cite{Dzuba99,Dzuba02}. The more accurate results from the 
later work are included in the Table. The Table also presents the
results of independent calculations by the St. Petersburg group~\cite{Porsev}.
This is the most detailed and accurate calculations which include
core-valence correlations and Breit corrections. The results of
both calculations agree within the declared accuracy. Accuracy
is high for lower states but deteriorates significantly higher in
the spectrum due to increasing configuration mixing and difficulty
in the achieving of basis saturation for as many as seven external 
electrons.

Lines of neutral iron has not been used in the analysis yet. 
However, the lines presented in Table~\ref{tb:fe} are observed
in the quasar absorption spectra and are planed to be included
into analysis. Therefore we have recently calculated the
$q$ coefficients for them~\cite{Dzuba08}. It turned out that
the values of $q$ are large and for some lines they are even
larger than for Fe~II. This is due to strong configuration mixing
which makes the transitions to be effectively two-electron
transitions (see Ref.~\cite{Dzuba08} for details). This
makes Fe~I a good candidate for the inclusion into the analysis.

The results for other atoms are presented in Table~\ref{tb:other}. The calculations 
were carried out in works of Refs.~\cite{Dzuba99,Dzuba01,Dzuba02,Berengut04,Berengut05,
Berengut06}. See these works for more detailed information. Here we present
only the most commonly used lines and the latest more accurate calculations.

\section{Future directions}

The analysis of quasar absorption spectra has already produced very important
results putting strong constrains on possible space-time variation
of the fine structure constants $\alpha$ and hinting that $\alpha$
might be smaller in early epoch. However the results are controversial 
and more work is needed to prove or dismiss the contradicting claims.
Resolving the disagreement between the analysis of the Keck and VLT 
data seems to be the most important thing at the moment. This should
involve careful cross re-analysis of the data by independent groups
of experts. 

It is also important to include more data into the analysis
to improve statistics significantly. For example, having sufficient  
statistics for different lines with different shifts would allow
to compare the results which come from the analysis of each line
separately and therefore exclude many possible systematics. 
Also, improving statistics may allow to study variation of the fine 
structure constant as a function of the red-shift parameter $z$ or 
as a function of the position in the sky (space variation).

There are probably many more lines observed in quasar absorption spectra
but still not used in the analysis. Inclusion of this data may also add
important information about $\alpha$ variation. For example, calculations
for lines of Fe~I which are available for the analysis reveal that the
frequencies shifts for these lines due to change of $\alpha$ are large
and significantly different for different lines. Therefore, if corresponding
frequency shifts are observed it would be hard to attribute them to
anything but variation of $\alpha$.

Another example is the inclusion of the weak M1 or E2 transitions between
the states of the same fine structure multiplet~\cite{Kozlov}.
It was argued that the anomalies in the fine structure intervals 
lead to enhancement of the sensitivity of the transition frequencies
to variation of the fine structure constant. This idea is similar
to what what suggested before for laboratory experiments~\cite{FSA}.

It is also important to include molecular lines into analysis. This
would allow to study variation of the ratio of the electron to proton
mass ($\mu$). Most of grand unification models suggest that the variation
of mass ratio and $\alpha$ are related and mass ratio is changing 
faster than $\alpha$. Therefore it might be easier to find manifestations
of variation of mass ratio. There are already published results of
such analysis performed by different groups~\cite{mole1,mole2} but
as for the case of $\alpha$ the results are controversial but
the other way around. Those who claim variation of $\alpha$ see 
no variation of $\mu$ and vise versa. Here again more study is needed
and there are some new interesting suggestions (see, e.g.~\cite{Flambaum07}).
 
\section{Conclusion}

The many-multiplet method which is based on the analysis of the frequencies
of strong electric dipole transitions in atoms and ions found in gas
clouds intersecting the sight line from Earth to distant quasars
prove to be a useful tool for the search of the variation of the
fine structure constant in quasar absorption spectra. The method
relies on atomic calculations to reveal the dependence of
atomic frequencies on the fine structure constant. Critical
compilation of the all relevant calculations performed by our
and some other groups and presented in this paper can serve as a 
reference point for future analysis. The results of the analysis
so far is controversial. The analysis of the data from Keck telescope
in Hawaii indicate that $\alpha$ might be smaller in early Universe
while similar analysis of the data from the VLT telescope in Chile
performed by different groups of researchers gives null result.
Both analysis use the same MM method and the same atomic
calculations. This means the the reason for disagreement is
probably not relevant to atomic calculations and is rather
in the data or the analysis. Recent re-analysis of the VLT
data performed by the authors of the analysis of the Keck 
data pointed to some serious problems in the method used by
the other group. Further revision of the VLT data is needed to
resolve all problems in the analysis and disagreement in the results.

\section*{Acknowledgment}

The work was supported in part by the Australia Research Council.

\end{document}